\documentclass[letterpaper]{article}
\usepackage{amsmath,amssymb,graphicx,enumerate,footnote}
\usepackage{verbatim,graphicx,fullpage,url}
\usepackage{graphicx,color,verbatim,enumerate}
\usepackage[firstpage]{draftwatermark}
\SetWatermarkText{PREPRINT}
\SetWatermarkLightness{0.9}
\SetWatermarkScale{4}
\begin{document}

\title{Is That Twitter Hashtag Worth Reading}
\author{Anusha A and Sanjay Singh \thanks{Sanjay Singh is with the Department of Information and Communication Technology, Manipal Institute of Technology, Manipal University, Manipal-576104, INDIA, E-mail: sanjay.singh@manipal.edu}}
%
\date{}
\maketitle
\begin{abstract}
Online social media such as Twitter, Facebook, Wikis and Linkedin have made a great impact on the way we consume information in our day to day life. Now it has become increasingly important that we come across appropriate content from the social media to avoid information explosion. In case of Twitter, popular information can be tracked using hashtags. Studying the characteristics of tweets containing hashtags becomes important for a number of tasks, such as breaking news detection, personalized message recommendation, friends recommendation, and sentiment analysis among others. 
In this paper, we have analyzed  Twitter data based on trending hashtags, which is widely used nowadays. We have used event based hashtags to know users' thoughts on those events and to decide whether the rest of the users might find it interesting or not.  We have used topic modeling, which reveals the hidden thematic structure of the documents (tweets in this case) in addition to sentiment analysis in exploring and summarizing the content of the documents. A technique to find the interestingness of event based twitter hashtag and the associated sentiment has been proposed. The proposed technique helps twitter follower to read, relevant and interesting hashtag.
\end{abstract}

%
%

\section{Introduction}
The rapid growth in the popularity of social networking and microblogging has lead to a new way of finding and broadcasting information in the past decade. The websites like Facebook, Twitter, Linkedin, Stack Overflow, and Stack Exchange, etc. have become the go to sites when people need any platform to communicate, broadcast their thoughts, publicity for an upcoming product or even a new app that they have developed. These websites bridge the gap between you and the rest of world in just one click.
\par
Twitter is noted to be the third most popular of such social networking sites incorporated on April 19, 2007 \cite{SMT}. It is one of the social networking platforms which has about 288 million active users generating 500 million tweets per day. Each tweet comprise of text up to 140 characters. Thousands of people advertise their recruiting services, consultancy, retail stores by using Twitter. The Internet users like this because it is less probing and available on mobile platform as well. These websites are increasingly used for communicating breaking news, eyewitness accounts and organizing large groups of people using hashtag \cite{hash} features in Twitter. Users of Twitter have become accustomed to getting regular updates on trending events, both of personal and global value. For instance, twitter was used to propagate information in real-time in many crisis situations such as the results of the Iran election \cite{TR}, the tsunami in Samoa \cite{Tsu} and more recently earthquake in Nepal \cite{Nepal}. 
\par
Many organizations and celebrities use their Twitter accounts to connect to customers and fans and propagate their thoughts. Twitter is depicted as a blend of instant messaging, microblogging, and knowledge source. Twitter is a way to connect with a person or a topic and decide for yourself if you like it. With such a progress in social networking, by analyzing textual data obtained from any of these sites which represent ideas, thoughts and communication between the users, it is possible to obtain an understanding of needs and concerns of the users that provides valuable information for academic, marketing and policy-making. 
\par 
Since there are about 500 million tweets generated every day and very vast number of users look for interesting tweets among those tweets, there is a need for a mechanism to find such interesting tweets using an unbiased method. It should not depend on author popularity and consider only the content of tweets. Every tweet generated by a user can be retweeted by other users those who are following the author of that tweet. Until recently, "retweet" count of a tweet was used as measure of popularity of that tweet regardless of topic and content of the tweet. However, retweet count depends mainly on the author popularity and not the content popularity of the tweet. More the followers an author has, the more publicity the tweet could obtain. In order to find the tweet that is of interest to large audience we need to consider more than just retweet count. Hence, we need to analyze the short text message that usually includes noise. The text messages from twitter needs to identify the interesting tweets among them by using an automatic method, so that it can be of use to wider range of audience.
\par
In order to find the interesting tweets, we have used topic modeling \cite{TopicModl} based on Latent Dirichlet Allocation (LDA) \cite{Blei}.
This is followed by sentiment analysis \cite{SA} which helps to build a more human like system. The aim here is to understand the sentiment of each tweet using natural language processing. We have utilized the 'hashtag' to extract tweets and find their interestingness and sentiment associated with that tweet.

\section{Related Work}
Topic models \cite{BleiTM} are powerful tools to identify latent text patterns in the content. It has been applied in a wide range of areas including recent work on Twitter \cite{2014}. Social media differs from some standard text domain (e.g., citation network, web pages) where topic models are usually utilized in a number of ways. One important fact is that there exists many dimensions in social media that we usually want to consider them simultaneously. Many studies have provided insights into social media. Kwak et al. \cite{Kwak:2010:TSN:1772690.1772751} studied Twitter's structure by investigating various Twitter features. Recently, many works \cite{Castillo:2011:ICT:1963405.1963500} \cite{Duan:2010:ESL:1873781.1873815} have focused mainly on analyzing or obtaining valuable information, such as influential users and posts on Twitter, from a large amount of social data. 
\par
Most of the existing approaches have considered retweet counts as a measure of popularity, influence, and interestingness, and presented classifiers that predicted whether and how often new tweets will be retweeted in the future. They exploited many features of twitter, such as textual data, author's information, and propagation information. Although the overall retweet count indicates a tweet's popularity, this may apply only to the followers of the tweet's author.
\par
Twitter not only has textual data but also has related data, such as follower and retweet links, which enable us to construct a network structure. The link-based approaches applied a variant of the link analysis algorithm \cite{Romero:2011} to a designed link structure in order to find interesting messages. However, the link structure requires a large volume of linking data to be analyzed and constructed and cannot be updated effectively when new tweets are generated. 
\par
Alonso et al.\cite{Alonso}, used crowdsourcing to categorize a set of tweets as interesting or uninteresting and reported that the presence of a URL link is a single, highly effective feature for selecting interesting tweets with more than 80 \% accuracy. This simple rule, however, may incorrectly categorize an uninteresting tweet (i.e., an uninteresting tweet contains links to meaningless pictures, videos, and advertisements)as interesting. Lauw et.al \cite{Luaw2010} suggested several features to identify interesting tweets but did not experimentally validate them. For user recommendation, Armentano et.al \cite{Armentano:2013:FRB:2533319.2533398} examined the topology of followers/followees network and identified the relevant users using social relation factors. They conducted not only topology-based profiling but also content-based profiling to find semantically similar users. 
\par 
In social media, semantic analysis and topic modeling are widely used to understand textual data and can facilitate many applications such as user interest modeling \cite{Pennacchiotti:2011:ITM:1963192.1963244}, sentiment analysis \cite{Lin:2009:JSM:1645953.1646003}, content filtering \cite{Duan:2013:WOT:2500516.2500657}, and event tracking etc. Zhao et al. \cite{Zhao:2011} analyzed the topical differences between Twitter and traditional media using TwitterLDA for investigating short messages. Wang and McCallum \cite{Wang:2006:TOT:1150402.1150450} and Kawamae \cite{Kawamae:2011:TAM:1935826.1935880} conducted topic modeling of temporally-sequenced documents in Twitter and tried to model the topics continuously over time. However, in our approach LDA considers the mixtures of latent topics as a trend based on hashtags and is designed to learn changes in topic distributions, while other works focus on learning topic shifts based on word distributions. 
\par
Chen et al.\cite{Chen:2010:STE:1753326.1753503} focused on recommending URLs posted in tweets using various combinations of topic relevance and social graph information. The model by Ramage et al.\cite{ram:2010} is an unsupervised learning method with relative importance of latent topics. In 2014, Min-Chul Yang and Hae-Chang Rim \cite{2014} proposed a better model based on Ramage's work to find the interesting tweets, but it does not include sentiment polarity analysis.
\begin{figure}[bpht!]
	\begin{center}
		\includegraphics[height=6.2cm, width=7cm]{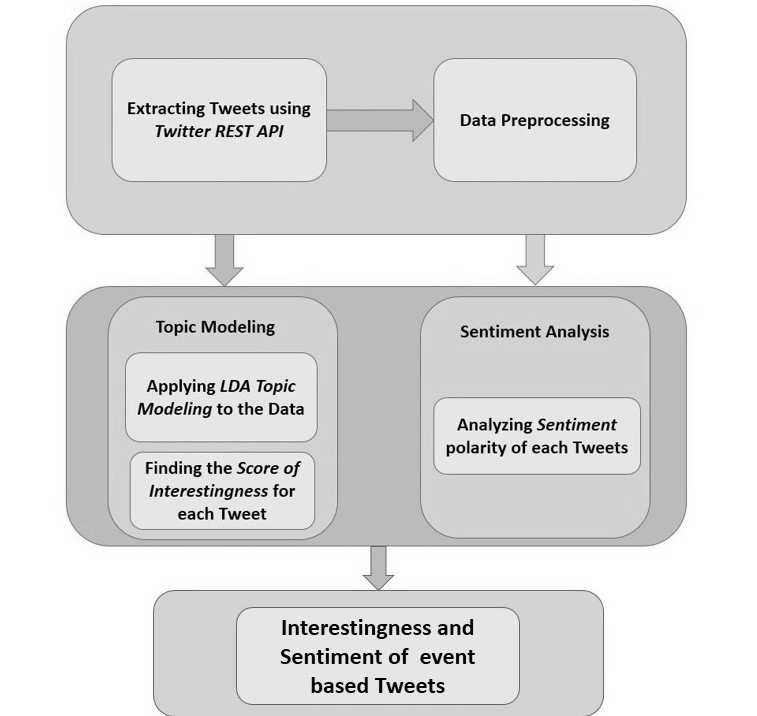}
		\caption{Block Diagram of Proposed System}
		\label{fig:f1}
	\end{center}
\end{figure}

\section{Proposed Method}
We collected tweets based on hashtags related to a specific event, ICC World Cup 2015, along with the tweets that was not generated by any particular event. Yang and Rim \cite{2014} used a time stamp approach in their work to find interesting tweets that belong to a specific time period, where as we have used hashtag feature that is already available in twitter to eliminate the need of time stamp and find the interestingness and sentiment of tweets generated using hashtags that are related to a specific event that has occurred or occurring at present.
\par 
Figure \ref{fig:f1} shows the block diagram of our proposed method. Once the tweets are extracted generally as well as based on hashtags, it needs to be preprocessed by removing stop words, discarding tweets with non English words, special characters, etc. Every individual tweet is considered as a document. Then such tweets are analyzed to find their interestingness and sentiment polarity. A higher score to interesting tweet is assigned and their polarity is found. The major concept used by Yang and Rim \cite{2014} is \textit{interestingness}. It indicates the number of audience that might find this tweet interesting to them without the consideration of author and author's popularity. 
\par
To find the interestingness we need to extract latent topics that tweets belong to. We have used LDA \cite{Blei} to infer latent topics that the tweets we have collected belongs to. LDA considers a document as a bag of words and not the context of the words that appearing in documents. Since all topics do not contribute to be a part of interesting tweet, so we find the weight as well as spatial entropy for each topic. Tweets that belong to the topics with higher weight and less spatial entropy is considered to be more interesting.
\par
\begin{figure}[bpht!]
	\begin{center}
		\includegraphics[height=3.2cm, width=6cm]{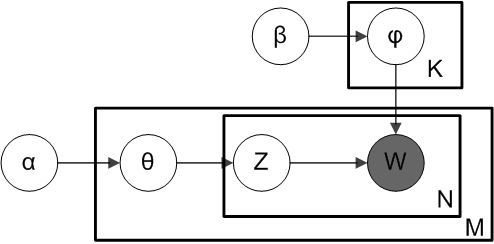}
		\caption{Plate Notation of LDA \cite{plate}}
		\label{fig:f2}
	\end{center}
\end{figure}

 
LDA generates the latent topics for the given set of documents and two probability distributions. First is the word distribution in each topic, and second is distribution of topics in each document. These two distributions are used to find the spatial entropy and integrity of each topic. The plate notation of LDA is given in Fig. \ref{fig:f2}.

In Fig.\ref{fig:f2} $N$ represents the number of words belonging to $K$ number of topics, being distributed over $M$ number of documents. Each document can belong to one or more topics and each word will have its own probability in topics. The $\alpha$ and $\beta$ are parameters that depict the probability of words and topics respectively that has not yet occurred in the data set, but may appear in future. We have set their value to be 0.01 and it can be set to any value between 0 and 1. Here $\theta$ is the probability of topic $z$ in the given document and $\varphi$ is the probability of word $w$ in the given topic. We have used The Stanford Topic Modeling Toolbox (TMT) \cite{TMT} to perform analysis on our data set. 
\par 
In order to find weight of the topics, we have to find integrity and spatial entropy of each topic using the distribution obtained after applying LDA on our data set.  

\subsection{Tweet Scoring Measures}
Each tweet is evaluated for its interestingness based on the topics it belongs to. If a tweet belongs to topics that are having higher weights, it is said to be more interesting. Note that the weight of each topic represents how important that topic is. If the topic contains words which are not standard words or if the topic is found to be present in too many documents then such topic is considered as a noisy topic. However, to find weight of each topic, first we need to calculate its integrity and spatial entropy. 

\subsubsection{Integrity}
Integrity of topic considers that every topic may not be useful to analyze the data that is collected. Each topic has its own distribution of words, and these words are used to determine the integrity of the topic. A lexical dictionary that has all English words and popular non-dictionary words is built manually. It contains more than 0.1 million words. The \textit{Integrity of Topic (I)} is measured as follows \cite{2014}:
\begin{equation}
I(t) = \sum_{w \in W} p(w|t)L(w)
\end{equation}
where $p(w|t)$ is the probability of word $w$ in the topic $t$ and $L(w)$ is 1 if the word is present in the lexical dictionary else 0. 

\subsubsection{Spatial Entropy}
Spatial Entropy depicts the distinction of topics based on topic distribution. Most meaningful topics are considered to be related to smaller number of documents. If a topic is found to be closely related to too many documents, then that topic is considered to be noisy or general. \textit{Spatial Entropy} is given by \cite{2014}:
\begin{equation}
S(t) = - \sum_{d\in D} p(d|t)\log p(d|t)
\end{equation}

where document $d$ is a single tweet and $p(d|t)$ is the probability of document $d$ given topic $t$. The values of $p(d|t)$ is calculated based on Bayesian inference in LDA \cite{Blei} . 
The \textit{Integrity} and \textit{Spatial Entropy} are normalized using 
\begin{equation}
 \tilde{x} = z(x) = \frac{x-\mu}{\sigma} 
\end{equation}
 
where $x$ is variable, $\mu$ is average of variables and $\sigma$ is standard deviation of variables.

\subsubsection{Weight of Topics}
Weight of topics is computed based on the normalized \textit{Integrity} and \textit{Spatial Entropy}. It is given by:
\begin{equation}
 W(t) = \tilde{I}(t)-\tilde{S}(t)
\end{equation}
 
The entropy value is subtracted from integrity value because lower the entropy higher the importance of the topic. The topic gets a low weight if it is found to be noisy or meaningless.
 
\subsubsection{Scoring Tweets}
 Scoring the interestingness of tweets is done by assigning relative topic score and probability of that topic in a particular tweet. The scoring function for tweet $t$ is given by
 \begin{equation}
 Score(t) = \sum_{t\in T} W(t)p(t|d)
 \end{equation} 
 
 where $W(t)$ is the calculated weight of topic $t$ and $p(t|d)$ is the topic distribution for document $d$. Eventually, the document which covers topics with higher weight gets higher score of interestingness. The tweet with the higher score is considered as the more interesting than the tweets that have lower scores. 
 
\subsection{Sentiment Polarity Analysis}
Sentiment polarity analysis is a natural language processing task which returns the polarity of each documents as a float value within the range [-1.0, 1.0]. The first value represent the positive polarity and second value represents the negative polarity. There are few NLTK corpora \cite{nltk} which is considered for training data and their analysis is considered in deciding the sentiment polarity of the documents that we provide. If the value of positive polarity is higher than the value of negative polarity, then the tweet is said to convey a positive sentiment and vice-versa. 

\section{Results and Discussion}
We have used \textit{Twitter REST API} \cite{REST} to collect tweets. The experimental results of our method to evaluate the interestingness and sentiment of tweets  is given in this section.

 \begin{figure}[bpht!]
   	\begin{center}
   		\includegraphics[height=4.5cm, width= 8.2cm]{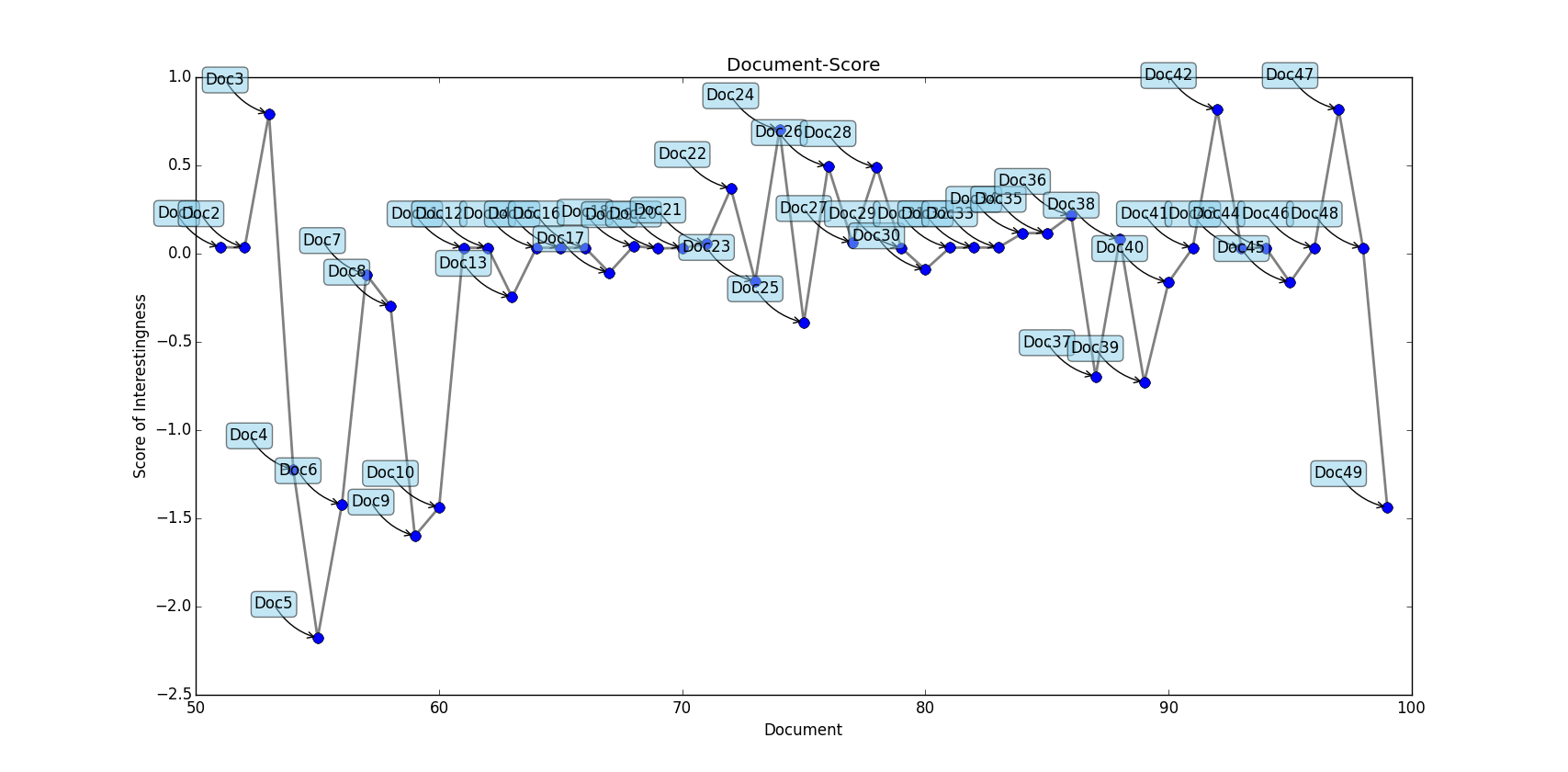}
   		\caption{Document-Score Plot}
   		\label{fig:f3}
   	\end{center}
   \end{figure} 
  
\begin{table}[bpht!]
\caption{Tweets and their Interestingness Score}
\label{tab:t1}
\begin{center}
\begin{tabular}{|p {5.5cm} | p{1.5cm} |}
 \hline\hline
 \textbf{Tweet} & \textbf{Score} \\
 \hline \hline
 Two host nations then just like the last time. Best 2 teams in the tournament fighting for $d$ title.  & 1.83701 \\
 
 \hline
 No interest in ICCWC.  & -0.65715 \\
 \hline
 Three years back on this day we won the ICCWC for the second time in ODIs. Cheers to the victory. & 2.07052 \\ 
 \hline
 Australia Cricket Champs Again, Congrats & 2.48527 \\
 \hline
 Injured Australian batsman Phillip Hughes dies& -0.8024191\\ 
 \hline
 Miss coming home from work to the World Cup  & 0.03527\\ 
 \hline  
 Loosing in cricket always hurts but take a bow, we have played some brilliant cricket, especially bowling @bcci & 1.40191\\ 
 \hline 
 Has Virat Kohli taken the responsibility to make them winner? & 0.043229\\ 
  \hline\hline
 \end{tabular} 
 \end{center}
\end{table}

\par
Interesting tweet classification task is completely content driven and the author's popularity does not influence this scoring mechanism in any way. For our method as tweets crawled were not more than 5K, we set the number of topics to be 15 which is based on our emprirical experiments and performed 1000 iterations of \textit{variational inference} in TMT. 
\par 
For sake of presentation the score of 50 such tweets related to ICC World Cup is shown in Fig.\ref{fig:f3}. Table \ref{tab:t1} lists eight tweets and their corresponding interestingness score. Among the eight tweets fourth is the most interesting tweet and fifth is found to be the least interesting tweet. We have considered each tweet as interesting tweet if the score obtained is above 1. Since the number of topics considered for our dataset is low, for a relatively low number of topics present in the dataset, the score threshold is also fixed to a lower value. However, depending upon the number of topics threshold for interestingness may vary accordingly.
\par 
We have collected tweets based on hashtags 'ICCWC' and 'CWC15' to analyze the sentiment based on the event of \textit{ICC World Cup 2015}. Their sentiment polarity along with the rest of the tweets were calculated, Table \ref{tab:t2} lists few such tweets and their sentiment polarity.
\begin{table}[bpht!]
\caption{Tweets and their Sentiment Polarity}
\label{tab:t2}
\begin{center}
\begin{tabular}[c]{|p{4cm} | p{1.5cm} | p{1.5cm} |}
 \hline
 \textbf{Tweet} & \textbf{Positive Polarity} & \textbf{Negative Polarity} \\
 \hline \hline
 @skysportnz thank you so much for securing the rights of @IPL. It was so boring after iccwc. Gonna be fascinating 6 weeks again & 0.9522 & 0.04779 \\
 \hline
 No interest in ICCWC. & 0.4573 & 0.5426 \\
 \hline
  Australia Cricket Champs Again, Congrats & 0.9171 & 0.0828 \\
 \hline
 Dhoni's innings was completely bizarre, what was his plan??? AUSvIND ICCWC  & 0.4252 & 0.5747 \\
 \hline
 Scotland vs Sri Lanka iccwc & 0.5 & 0.5 \\
 \hline
 \end{tabular} 
 \end{center}
\end{table}

\begin{figure}[bpht!]
   	\begin{center}
   		\includegraphics[height=4.5cm, width= 8.2cm]{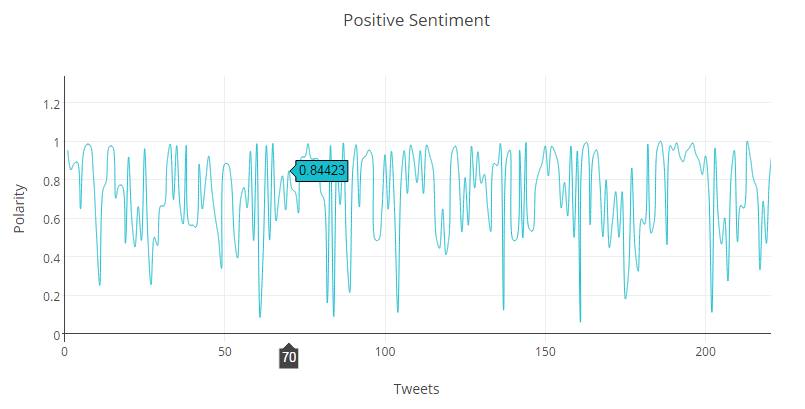}
   		\caption{Sentiment Variation on ICCWC}
   		\label{fig:f4}
   	\end{center}
   \end{figure}

\begin{figure}[bpht!]
   	\begin{center}
   		\includegraphics[height=4.5cm, width= 8.2cm]{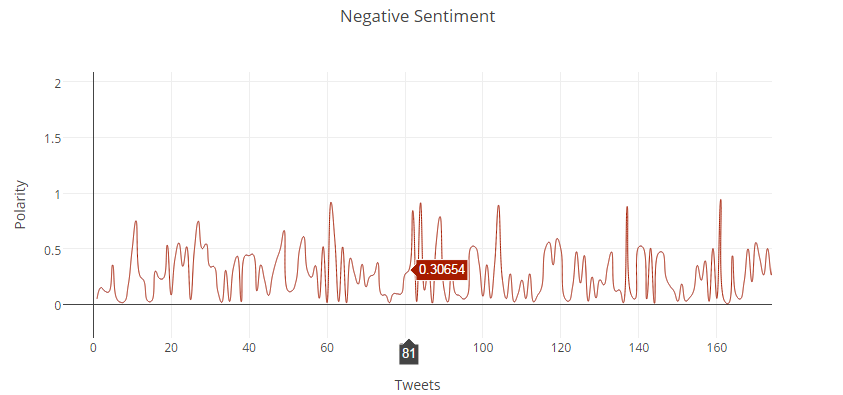}
   		\caption{Sentiment Variation on ICCWC}
   		\label{fig:f5}
   	\end{center}
   \end{figure}

Figure \ref{fig:f4} and Fig.\ref{fig:f5} depicts the sentiment variation of few tweets related to the event of ICC World Cup 2015. Figure \ref{fig:f4} shows the positive sentiment variation and Fig.\ref{fig:f5} shows the variation in negative sentiment.
 
   
\begin{figure}[bpht!]
   	\begin{center}
   		\includegraphics[height=6cm, width= 9.2cm]{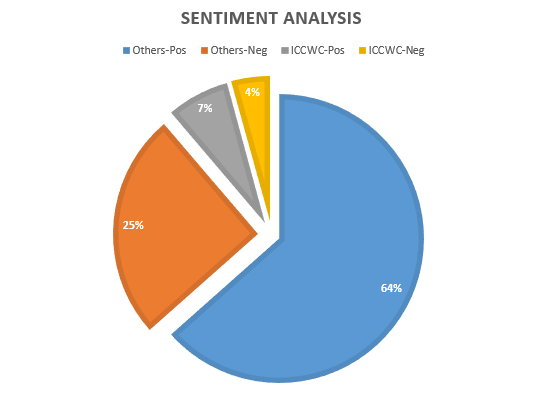}
   		\caption{Sentiment Analysis}
   		\label{fig:f6}
   	\end{center}
   \end{figure}

As we mentioned earlier the hashtags related to the event of ICC World Cup was included to extract tweets and the result of this sentiment analysis is shown in Fig.\ref{fig:f6}.  From Fig.\ref{fig:f6} we observed that the most of the tweets were not completely positive or negative. For example, when New Zealand won against South Africa in semifinals, people felt bad for South Africa but appreciated New Zealand's game as well. This generates almost equal values of both positive and negative polarity to such tweets. 

Every user has their own opinion about an event occurring. The tweets belong to various topics, events may convey different emotions: positive or negative, which needs to be analyzed before deciding if the tweet is interesting or not. The proposed method can handle numerous latent topics that can be found in large set of data. The probability distribution obtained after applying LDA is reliable and gives expected results, which can be followed by sentiment analysis. It would be good if the dataset is sufficiently large to find proper latent topics. Our sentiment analysis is done against a vast annotated corpora and tend to take a longer time to analyze a large dataset.

\section{Conclusion}\label{conc}
In this paper, we have proposed an effective approach to evaluate the interestingness of tweets based only on the textual contents and find what sentiment they convey. Topic modeling using LDA was applied to discover the latent topics and then find the weight of each topic, which depends on their integrity and spatial entropy. Higher integrity and lower spatial entropy contributes to a higher weight for the corresponding topic. The tweets that are distributed over topics with higher weight is considered as more interesting and if the tweet belongs to topics with lower weight it is said to be less interesting.
\par 
In addition to this, we performed sentiment analysis on tweets to find the sentiment polarity of each tweet and to find the variation of sentiment with time and with respect to an event that is occurring. Our approach can be used to find the interestingness and sentiment polarity of tweets related to an event or in general. \par 
In future, we can think of applying this method to a real time system to suggest tweets that belong to specific topic and depicts the desired sentiment. 
Since the hashtags are commonly used feature in most of the social media and not just in Twitter, our approach can be used to analyze any platform's textual data that is related to an event by hashtag.

\bibliographystyle{IEEEtran}
\bibliography{myref}

\end{document}